\documentclass[epj,referee]{svjour}

\usepackage{graphics}

\usepackage{rotate}
\usepackage{epsfig}
\usepackage{psfrag}

\newcommand \be  {\begin{equation}}
\newcommand \bea {\begin{eqnarray}\nonumber}
\newcommand \ee  {\end{equation}}
\newcommand \eea {\end{eqnarray}}

\begin{document}
%\authorrunning
%\titlerunning
\title{Spatial correlations in vote statistics: a diffusive field model for decision-making}

\author{Christian Borghesi\inst{1}\thanks{christian.borghesi@cea.fr} \and Jean-Philippe Bouchaud\inst{2}\thanks{jean-philippe.bouchaud@cea.fr}}
\institute{Service de Physique de l'\'Etat Condens\'e, Orme des Merisiers, CEA Saclay, 91191 Gif-sur-Yvette, France \and Science \& Finance, Capital Fund Management, 6-8 Bd Haussmann, 75009 Paris, France}

\date{\today}% The correct dates will be entered by Springer

\abstract{
We study the statistics of turnout rates and results of the French elections since 1992. We find that the distribution of turnout rates across towns is surprisingly stable over time. 
The spatial correlation of the turnout rates, or of the fraction of winning votes, is found to decay logarithmically with the distance between towns. 
Based on these empirical observations and on the analogy with a two-dimensional random diffusion equation, we propose that individual decisions can be rationalised in terms of 
an underlying ``cultural'' field, that locally biases the decision of the population of a given region, on top of an idiosyncratic, town-dependent field, with short range correlations. 
Using symmetry considerations and a set of plausible assumptions, we suggest that this cultural field obeys a random diffusion equation.
} %end of abstract

\titlerunning{Spatial correlations in vote statistics: a diffusive field model for decision-making}
\authorrunning{C. Borghesi \and J.-P. Bouchaud}
\maketitle

\tableofcontents

% *********************************************************

\section{Introduction}

Making decisions is an everyday necessity. In many cases, these decisions are of binary nature: to buy or not to buy a product or a
stock, to get vaccinated or not, to participate to a vote or not. Elections are in fact themselves often binary, like referendums, or second round
of presidential elections, etc. The final decision of each individual is the result of many factors: individual propensities, common factors
(such as prices, reliability of the vaccine, importance of the election, etc.), and, in many cases, the decisions of others play a major role as well. 
Whether we like it or not, imitation is deeply rooted in living species as a strategy for survival. We, as humans, are influenced by our kindred 
both at a primitive level (fear of being excluded from the group) and at a rational level (others may possess some information unknown to
us). Imitation can lead to collective effects like trends, fashions and bubbles that would be difficult to understand if we were insensitive 
to the behaviour of others. These imitation induced opinion shifts can be beneficial for a society as a whole (as in the case of vaccination), 
but can also be detrimental and lead to major catastrophes (crowd panic, financial crashes, economic crises, rise of extremist ideologies, etc.). 
Developing reliable models for these collective effects is therefore of primary importance, see~\cite{granovetter,schelling,diffusion,galam_galam,brock,helbing,collective_shift,nadal_dicrete_choices}. This requires, in particular, to garner quantitative empirical information about the imitation processes that may induce strong 
distortions in the final outcome (see for example~\cite{salganik_exp} for 
a precise experimental set up to measure these effects, and~\cite{of_songs} for a theoretical framework.)

In order to study the nature of these behavioural correlations, we have studied the space-resolved (town by town) results of the French elections since 1992.
To keep the model and the interpretation of the results simple, we restrict to binary choice situations, i.e, either the turnout rates for 
each election, or the results of yes/no referendums or second round of presidential elections. We analyse in details the statistics of these
outcomes, with special focus on the dependence of the results on the size of the cities, and on the spatial correlations between the different
results. While there might be some indications of direct imitation effects within towns, the structure of inter-town correlations strongly suggests the
existence of what we propose to call a `cultural' field, that evolves in time according to a noisy diffusion equation. This cultural field encodes local 
biases in intentions, convictions or propensities on a given subject, for example to vote or not to vote, or to vote left or right, to 
respect or not speed limitations, etc. etc.
These (subject specific) cultural fields transcend individuals while being shaped, shared, transported and transformed by them. 
Although the existence of such cultural fields has been anticipated by sociologists, political scientists and geographers (see \cite{Siegfried} for an early 
insight, and \cite{bussi_geo_electorale,web} for more recent discussions), we believe that our empirical 
results provide the first {\it quantitative} evidence for such a concept, and lay forward the possibility of a precise modelisation of its spatio-temporal
evolution. Let us emphasise an important 
conceptual point: these cultural fields should exist {\it independently of any election}, or any other 
occasions where a decision has to be made. These events provide an instantaneous snapshot of the opinion or of the behaviour of individuals, 
which are in part influenced by these fields, in a way that we will quantify below. The cultural field has a dynamics of its own, that we will model and 
elaborate on in section \ref{randif} below. 

Most of the empirical electoral studies previously reported in the physics literature deal with proportional voting from multiple choice lists, 
and investigate the distribution of votes; as in Brazil~\cite{costa_filho_scaling_vot,costa_filho_bresil_el2,lyra_bresil_el,bernardes_bresil_el}, 
in Brazil and India~\cite{gonzalez_bresil_inde_el}, in India and Canada~\cite{hit_is_born}, in Mexico~\cite{baez_mexiq_el,morales_mexiq_el}, 
in Indonesia~\cite{situngkir_indonesie_el}. A universal behaviour was reported in~\cite{fortunato_universality}. Statistical results of elections for the city mayor 
are studied in~\cite{araripe_plurality}, the typology of Russian elections in~\cite{sadovsky_russie_el}, correlations between electoral results and party members 
in Germany in~\cite{schneider_impact}, and statistics of votes per cabin for three Mexican elections in~\cite{hernandez_bvot_mexique}. 
Majority and Media effects were investigated for various countries in~\cite{growth_model_vote}. Lastly, the spread of Green Party in several states in the USA 
is analysed by means of epidemiological models in~\cite{third_party_epidemiological}, and data from a 
Finland election is confronted to a Transient Opinion Model~\cite{banisch}. 
See also \cite{blais,franklin} for Political Science studies of electoral participation.

The specific feature of the present work is the quantitative analysis of the {\it spatial correlations} of the voting patterns. 
We will first describe several striking empirical regularities in the French vote statistics (that we believe are not restricted to French
elections). We then turn to simple models that help putting these findings in context, and explain why the idea of a diffusive cultural 
field, which is the central proposal of this work, naturally accounts for some of our findings.

\section{Empirical regularities and spatial correlations}

\begin{table*}
\caption{Summary statistics of the elections studied in this paper. R.: referendum (1992: Maastricht treaty, 2000: reduction of the presidential mandate to 5 years, 2005: European 
constitutional treaty); E.: European parliament election; P.1: first round of a presidential election; 
P.2: second round of a presidential election. Mean value (mean), standard-deviation (sd), skewness (skew) and kurtosis (kurt) of logarithmic turnout rates $\tau_\alpha$ (left), and of 
logarithmic winning vote rates $\rho_\alpha$ (right). Data is for metropolitan France only.}
\label{telections}
\begin{tabular}{|c|c||c|c|c|c|c||c|c|c|c|c|}
\hline\noalign{\smallskip}
election & kind & $\frac{\sum_\alpha V_\alpha}{\sum_\alpha N_\alpha}$ & mean & sd & skew & kurt & $\frac{\sum_\alpha W_\alpha}{\sum_\alpha V_\alpha}$ & mean & sd & skew & kurt\\
\noalign{\smallskip}\hline\noalign{\smallskip}
1992-b & R. & 0.713 & 1.13 & 0.355 & 1.05 & 5.48 & 0.508 & -0.164 & 0.447 & -0.159 & 2.48\\
1994-m & E. & 0.539 & 0.358 & 0.398 & 0.837 & 9.23 & & & & &\\
1995-m & P.1 & 0.795 & 1.60 & 0.375 & 0.928 & 5.37 & & & & &\\
1995-b & P.2 & 0.805 & 1.72 & 0.398 & 1.35 & 5.54 & 0.525 & 0.187 & 0.524 & 0.357 & 2.71\\
1999-m & E. & 0.478 & 0.146 & 0.392 & 1.15 & 7.50 & & & & &\\
2000-b & R. & 0.308 & -0.626 & 0.377 & 0.858 & 8.27 & 0.729 & 0.874 & 0.498 & -0.116 & 2.77\\
2002-m & P.1 & 0.729 & 1.24 & 0.347 & 1.25 & 9.46 & & & & &\\
2002-b & P.2 & 0.810 & 1.67 & 0.367 & 1.26 & 6.40 & 0.820 & 1.48 & 0.521 & 0.776 & 2.26\\
2004-m & E. & 0.434 & -0.095 & 0.366 & 1.45 & 9.82 & & & & &\\
2005-b & R. & 0.711 & 1.13 & 0.351 & 1.58 & 12.0 & 0.550 & 0.377 & 0.443 & -0.021 & 1.37\\
2007-m & P.1 & 0.854 & 1.98 & 0.396 & 1.06 & 8.02 & & & & &\\
2007-b & P.2 & 0.853 & 1.99 & 0.394 & 1.22 & 5.28 & 0.533 & 0.257 & 0.487 & 0.174 & 2.31\\
2009-m & E. & 0.414 & -0.147 & 0.360 & 1.35 & 8.28 & & & & &\\
\noalign{\smallskip}\hline
\end{tabular}
\end{table*}

We have analysed the turnout rate of all French elections~\cite{data_fr} with national choices since 1992 (13 events). A subset of 6 elections offered a binary 
choice to voters: 3 referendums and 3 second round of presidential elections (See Table~\ref{telections} for details and summary statistics). The national results are broken 
down into $\simeq 36,000$ local results, corresponding to {\it communes} (towns), of various population sizes. For each voter $i$, we define $S_i=1$ to
correspond to participation to the vote or belonging to the majority vote, whereas $S_i=0$ corresponds to not participating, or belonging to the
minority vote. For a town $\alpha$, the total number of potential voters is $N_\alpha$; the total number of voters is 
$V_\alpha=\sum_{i=1}^{N_\alpha} S_i$, the turnout rate is $p_\alpha=V_\alpha/N_\alpha$ and the total number of winning votes is $W_\alpha$. 
We found convenient to work with logarithmic rates for participation or winning votes: $\tau_\alpha=\ln(V_\alpha/N_\alpha-V_\alpha)=\ln(p_\alpha/1-p_\alpha)$ 
and $\rho_\alpha=\ln(W_\alpha/V_\alpha-W_\alpha)$. Each 
{\it commune} is characterised by the spatial coordinates~\cite{ign} of its {\it mairie} (town-hall), $\vec R_\alpha$. The distance between two {\it communes}, $r_{\alpha\beta}$, is defined
as $r_{\alpha\beta}=|\vec R_\alpha - \vec R_\beta|$ (even if the presence of -- say -- mountains or rivers in between would make the actual travelling distance much 
longer).

%histo abst
\begin{figure}
  \includegraphics[width=8.5cm, height=6cm, clip=true]{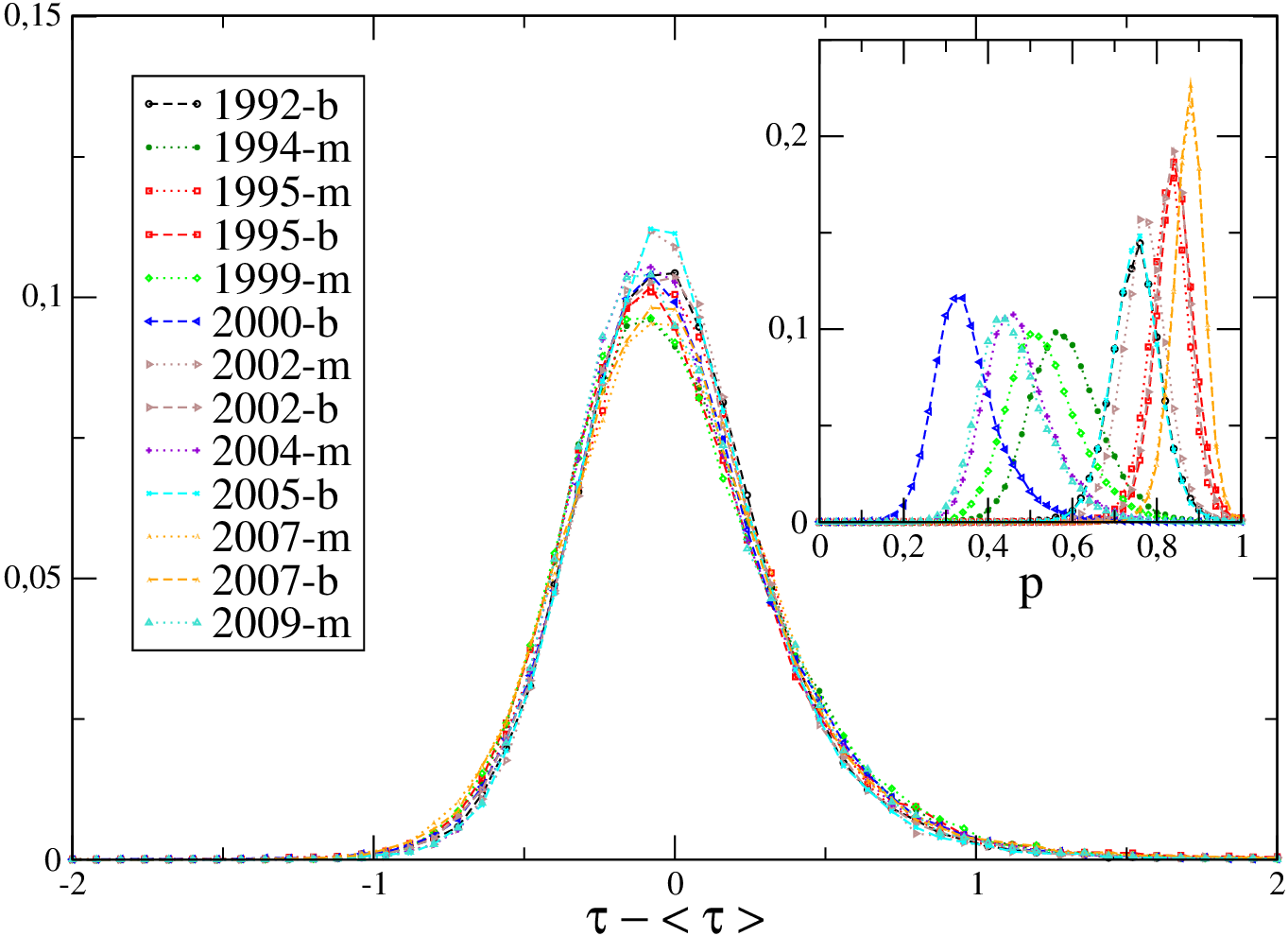}
  \caption{Distribution over \textit{communes} of $P(\tau - \langle \tau \rangle)$ for each election, where $\tau$ is the logarithmic turnout rate 
  and $\langle \tau \rangle$ its average over all \textit{communes}. The suffix -b means the election is of binary nature, while -m means multiple choice elections. 
  The inset shows the distributions of direct turnout rates, $p$, for each election.}
 \label{fhisto-abst}
\end{figure}

The issues at stake in all these elections are clearly very different, and so it is not a priori obvious that anything universal (across 
different elections) can be found in the statistics of votes. However, to our surprise, we found a number of empirical regularities that we now
detail, focussing first on turnout rates where these regularities are most robust. Very similar results 
are also found for winning votes -- see below. 

The simplest quantity to look at is the pdf of the logarithmic turnout rate $\tau$ over different {\it communes}, i.e. what is the probability 
$P(\tau) {\rm d}\tau$ that a given {\it commune}, irrespective of its size, has a turnout rate $\tau$ to within ${\rm d}\tau$. Although the average turnout 
rate $m=\langle \tau \rangle$ varies quite substantially between elections (see Table~\ref{telections}), the shape of the distribution of $\hat \tau = \tau 
- m$ is remarkably constant -- even without rescaling by the root mean square $\sigma$, see Fig.~\ref{fhisto-abst}. 
The first three cumulants of $P(\tau)$ are, within error bars, the same for all elections (see Table~\ref{telections}). In fact, 
a Kolmogorov-Smirnov test where one only allows for a relative shift of the distributions does not allow one to reject 
the hypothesis that $P(\hat\tau/\sigma)$ is indeed the same for all elections (except perhaps 2009 which gives marginal results). Note that the
distribution of $\tau$ is clearly non Gaussian, with significant positive skewness and kurtosis.

\begin{figure}
 \includegraphics[width=8.5cm, height=6cm, clip=true]{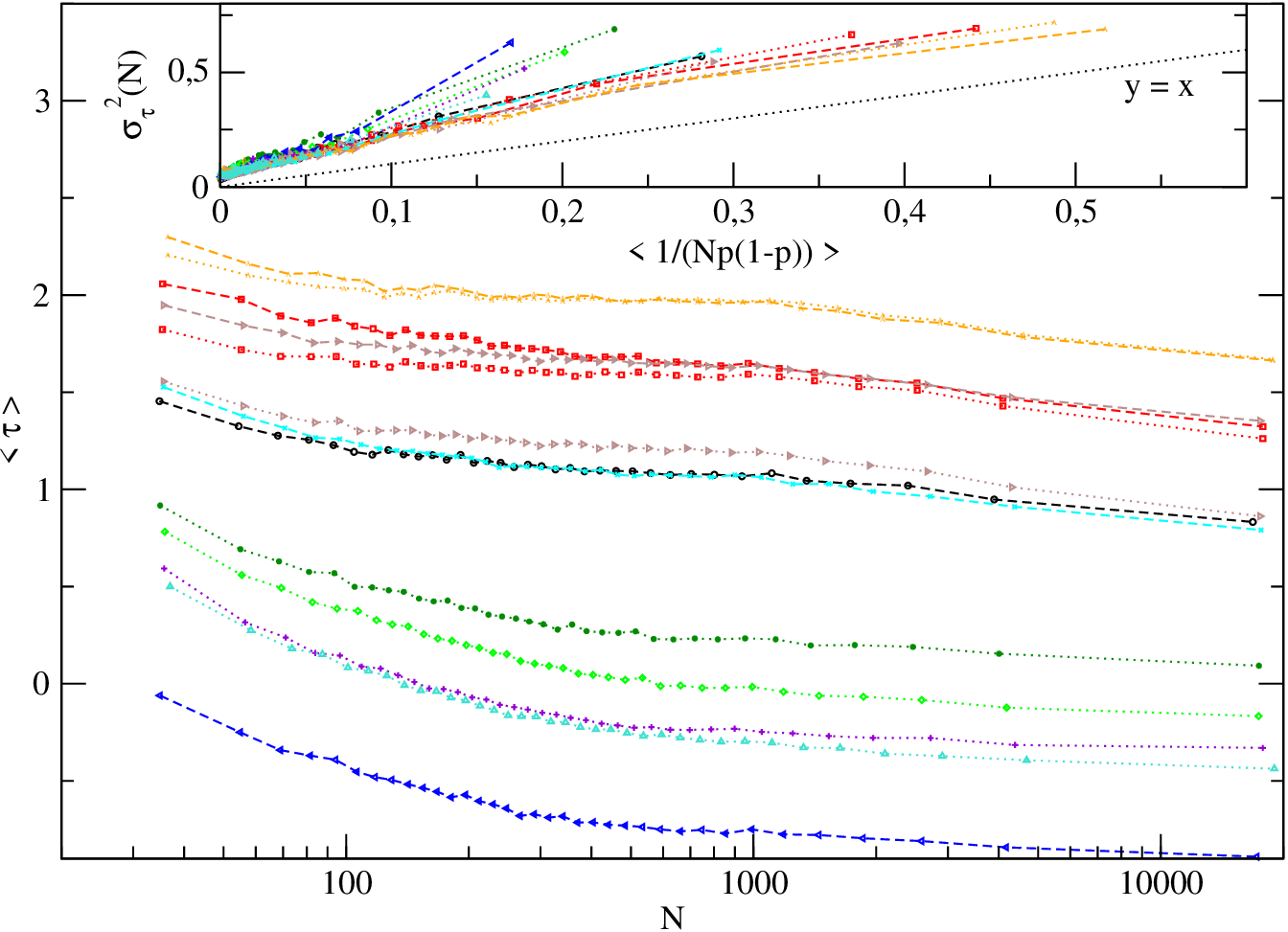}
  \caption{Mean (main figure) and variance (inset) of the conditional distribution $P(\tau|N)$. 
  These quantities are obtained as averages over bins with one thousand \textit{communes} of size $\approx N$.}
 \label{ftau-N}
\end{figure}

The statistics of $\tau$ does in fact depend on the {\it commune}-size $N$. We find that both the mean and variance of the conditional distribution $P(\tau|N)$ 
decrease with $N$ -- see Fig.~\ref{ftau-N}; in particular, small {\it communes} have a larger average turnout rate (this explains the positive skewness of 
$P(\tau)$ noted above~\cite{Borghesi}), but also a larger dispersion around the average. This is of course expected for a simple binomial process, 
which predicts that $\sigma_\tau^2(N) = 1/Np(1-p) + \Sigma^2$, where $p$ is the ($N$-dependent) average turnout rate, whereas $\Sigma^2$
describes the `true' variance of the turnout rate. The simplest assumption is that $\Sigma^2$ is $N$-independent, in which case the observed variance
varies significantly faster (by a factor $2$ or so) than the simple binomial prediction that assumes independent voters, see Fig.~\ref{ftau-N}, inset. 
A possible interpretation is that the votes of different individuals 
are not independent within the same {\it commune}, leading to an effectively smaller value of $N$ in the binomial. One can for example assume that within families, or groups of close
friends, the decision to vote or not to vote is exactly the same. If the size distribution of these groups of ``clones'' is $Q(s)$, then it is easy to show that for
$N$ large enough, these correlations in decision amount to replacing $N$ by $N \langle s \rangle_Q/\langle s^2 \rangle_Q \leq N$. An explicit example is $Q(s)=(1-z) z^{s-1}$
for $s=1,2,\dots$, which leads to an effective value of $N$ reduced by a factor $(1-z)/(1+z)$. In order to account for a factor of $2$ in the variation of $\sigma_\tau^2$
with $N$, we therefore need to choose $z \approx 1/3$. Since $z$ is the probability that the group is larger than $s=1$, this looks quite large. In fact, as we will discuss below, 
there is an alternative interpretation of the excess variance that does not rely on direct imitation.

%tau = f(N) + variance

Let us now turn to the spatial correlations between turnout rates, measured by the following correlation function:
\be
C_\tau(r) = \frac{\left.\left\langle (\tau_\alpha-m_\alpha)(\tau_\beta-m_\beta) \right\rangle\right|_{r_{\alpha\beta}=r}}{\langle (\tau_\alpha-m_\alpha)^2 \rangle},
\ee
where $m_\alpha$ is the average of $\tau$ over all {\it communes} with size in the same ``bin'' as that of $\alpha$, so as to remove systematic spatial correlations between town sizes. 
The central result of our empirical study is that for all elections, $C_\tau(r)$ is {\it long-range correlated}. It is found to decay as the logarithm of the distance (see Fig. 3, left): 
for $0 < r < L$, $C_\tau(r)=-\lambda^2 \ln(r/L)$, and $C_\tau(r > L) \approx 0$. Whereas the logarithmic slope $\lambda^2$ depends on the election and varies by a factor at most 
$2$ (between 1999 and 2007) 
the cut-off distance $L$ is, remarkably similar for all elections, with $L \approx 300$ km. In order to visualise more directly the correlations between neighbouring towns, 
we have also studied the relation between $\tau_\alpha$ and the average of $\tau_\beta$ over the 16 $\beta$ towns closest to $\alpha$. We find a very good linear regression 
between the two over the whole range of $\tau$'s, with a slope equal to unity (not shown, see \cite{Borghesi}). 

The same general picture holds for the statistics of winning votes, except that: (a) the distribution of the logarithmic winning rate $P(\rho)$ is much less universal across 
elections than $P(\tau)$. We have noticed in particular that the skewness of $P(\rho)$ varies significantly between elections, and changes sign between presidential elections, 
where it is positive, and referendums, where it is negative. Contrarily to $\langle \tau \rangle|_N$, $\langle \rho \rangle|_N$ does not show any systematic pattern, whereas $\sigma_\rho^2(N)$ is again 
significantly larger than the simple binomial prediction; (b) the spatial correlation of winning votes $C_\rho$ is also logarithmic with $r$ (see Fig. 3, right), although the 2002 result shows
a more pronounced curvature. The value of the cut-off distance $L$ is similar to the one reported above, whereas the slope $\lambda^2$ tends to be larger, except in the 2000 
election (see~\cite{Borghesi} for more details, and also for more statistical regularities in these elections.)

\section{Theoretical insights and threshold models}

The long-range, logarithmic dependence of the correlation functions $C_\tau(r)$ and $C_\rho(r)$ is the most striking finding of our study, in particular because it is 
strongly reminiscent of the behaviour of the correlation function of a free diffusion field in two dimensions. More precisely, let $\phi(\vec R,t)$ be a two-dimensional field that obeys the following
stochastic dynamical equation:
\be\label{EW}
\frac{\partial \phi(\vec R,t)}{\partial t} = D \Delta \phi(\vec R,t) + \eta(\vec R,t),
\ee
where $\Delta$ is the two-dimensional Laplacian, $D$ is a diffusion constant and $\eta$ a Langevin noise with zero mean, variance $\sigma_\eta^2$ and short-range correlations both in time and in space. 
It is well known that the equal-time correlation of $\phi$ is (in equilibrium) given by:
\be\label{log}
C_\phi(r) = \frac{\left \langle \phi(\vec r) \phi(0) \right \rangle}{\langle \phi(0)^2 \rangle} \approx - \Lambda^2 {\ln \frac{r}{L}},
\qquad \ell_c \ll r \ll L,
\ee
where $\ell_c$ is a short scale cut-off (for example the correlation length of the noise $\eta$) and $L$ is the linear size of the system. The behaviour of $C_\phi(r)$ for $r < \ell_c$ and the 
logarithmic slope $\Lambda^2$ depend on short scale details of the model, but not on the the diffusion constant $D$. The time to reach equilibrium, beyond which the above result holds, is 
$T_{eq} = L^2/D$. The logarithmic behaviour is a hallmark of 
the two-dimensional nature of the problem. The striking similarity between this prediction and our empirical findings is the motivation of the theoretical analysis that we present now.

%Correl spatiales
\begin{figure*}
\begin{center}
 \includegraphics[width=8.5cm, height=6cm, clip=true]{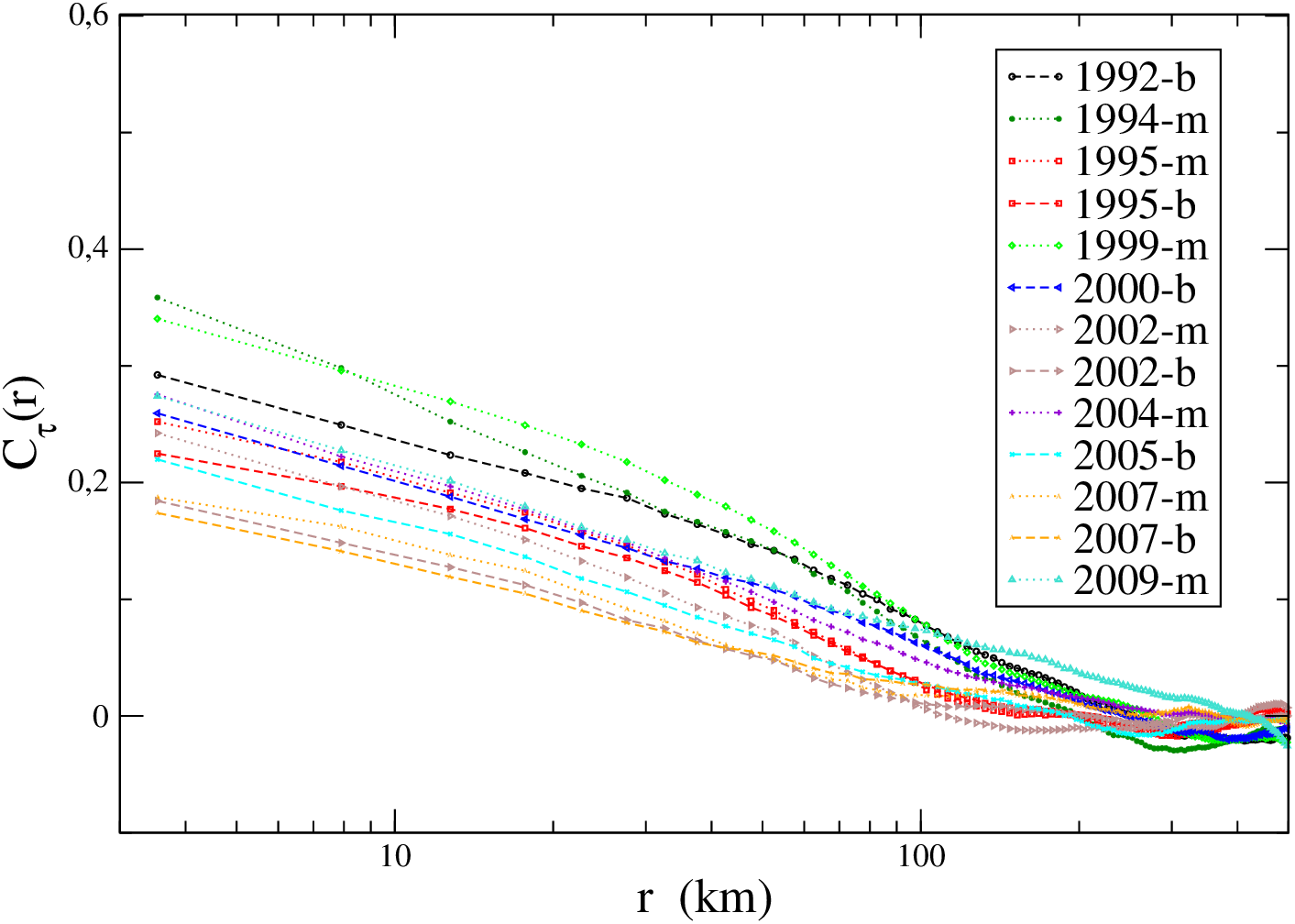}\hfill%, height=6cm
 \includegraphics[width=8.5cm, height=6cm, clip=true]{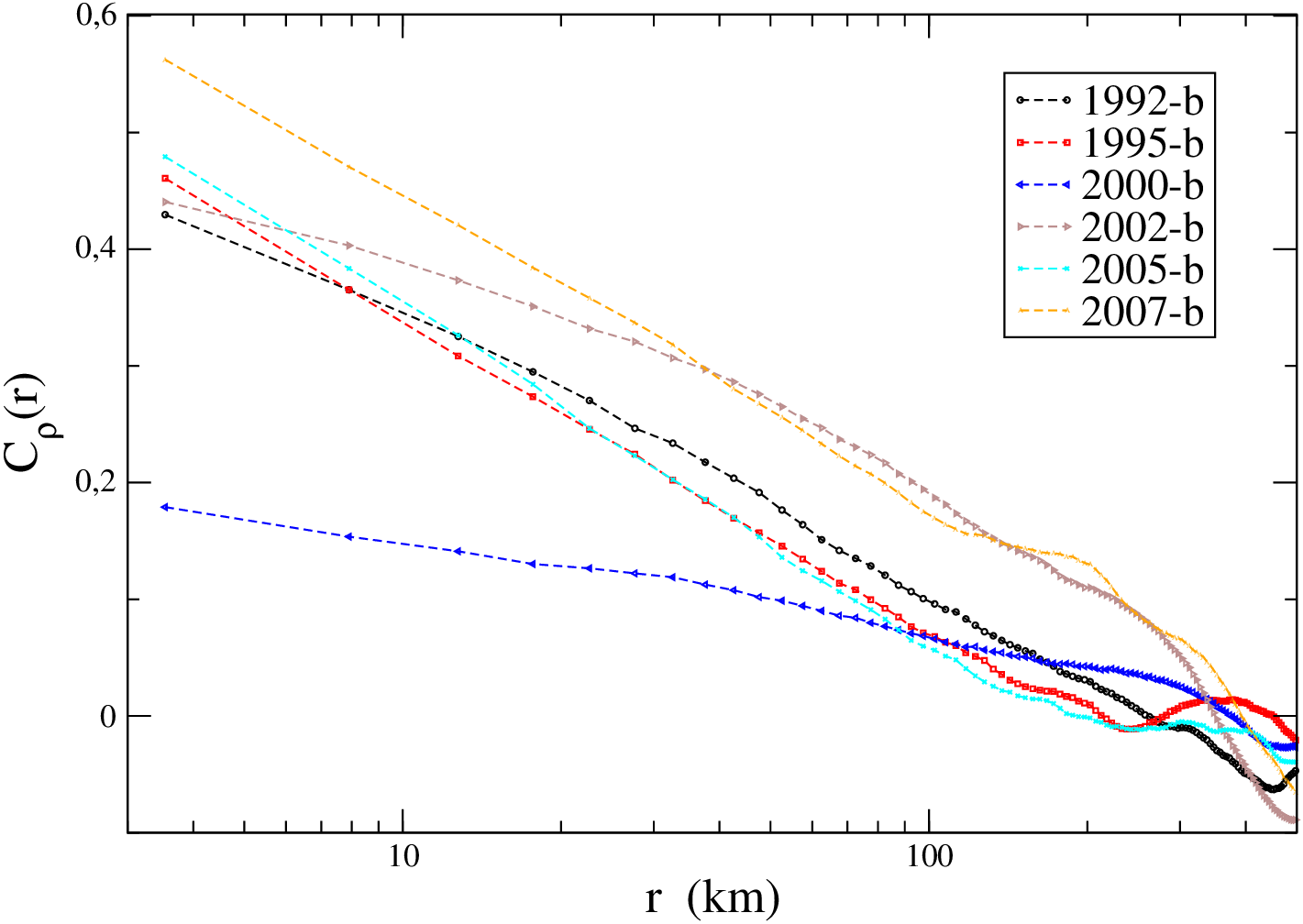}%, height=6cm
\end{center}
\caption{Spatial correlations $C_\tau(r)$ of logarithmic turnout rates (left) and $C_\rho(r)$ of winning votes (right), showing a logarithmic dependence on $r$ with a 
cut-off distance $L \approx 300$ km. 
The average logarithmic slope $\lambda^2$ is found to be $\approx 0.065 \pm 0.01$ for turnout rates, and $\approx 0.11 \pm 0.02$ for winning votes (removing the 2000
referendum on the reform of the presidential mandate, considered to be a very technical subject 
for which the turnout rate was exceptionally low, see Table 1.)}
\label{fcor-spatial}
\end{figure*}

There is quite a large literature on binary decision models (see \cite{granovetter,Anderson} for classical references, and \cite{brock,galam_galam,collective_shift,nadal_dicrete_choices,fortunato_stat_phys,stauffer_sociophys,bettencourt_epidemiological,schweitzer_brownian} for more recent contributions), although the spatial correlations that we want to include appear to be new. \footnote{We are aware of a ``spatial theory of political choices'' (see \cite{political}), but this is a misnomer: `spatial' in that case refers to a distance in the abstract space of convictions.} 
It is natural to think about these situations in terms of {\it thresholds} \cite{granovetter}: although the decision is binary, 
the process leading to the final choice is in fact continuous and reflects the individual motivations, propensities or utilities, that we will call the {\it intention} $\varphi_i$, 
where $i$ labels the individuals. 
When $\varphi$ exceeds a certain threshold $\Phi_c$, the decision $S$ is one way -- say $S=1$; when $\varphi$ is below this threshold, $S=0$. The process $\varphi$ 
is in general time dependent because individuals are influenced by a variety of factors: what they read, what they hear, what they see contribute to the way they understand a given situation
and react to it. 
Some of these influences are idiosyncratic, i.e. unique to each individual (for example, one may be ill on an election day, meaning that the propensity to go out and vote is very low), while others 
are common to people living in the same area (for example, the quality of public goods in a given town, or the unemployment rate, etc.). Finally, an important influencing factor is the 
decision $S_j$ made by others, if it is known before making one's own mind, or by the intentions $\varphi_j$ of others 
(see e.g.~\cite{granovetter,galam_galam,brock,collective_shift,nadal_dicrete_choices,salganik_exp,collective_attention}). 
For example, the decision to go see a movie, or to carrying on clapping at the end of concerts or shows, may well depend on what others have done in the past or are currently doing. 
In the first example, the number of people having already seen the movie feeds back on the motivation of 
those who have not yet seen it, through box-office charts or word of mouth. In the second example~\cite{collective_shift}, the amplitude of clapping generated by the rest of the crowd can be directly perceived 
and is an incentive to continue the applause. But in other situations where
the decision is taken simultaneously and the final result is only known {\it a posteriori} (like in elections), this direct imitation mechanism cannot be present -- although of course the 
influence of intentions is possible, and will be explicitly included in the model below.

The above discussion suggest a general decomposition of the individual intention field $\varphi_i$ into an 
idiosyncratic part, a `cultural' part and an imitation part. More formally, for an individual $i$ living in the vicinity of $\vec R$, the intention at time $t$ is written as: \footnote{Memory effects, or 
more complicated time dependent effects can be implemented along the lines of~\cite{marsili_ising_memory,collective_shift}.}
\be\label{conviction}
\varphi_i(t) = \epsilon_i(t) + \phi(\vec R,t) + \sum_{j} J_{ij} S_j(t-1);
\ee
where $\epsilon_i(t)$ is the instantaneous contribution to the intention that is specific to $i$, and $\phi(\vec R,t)$ is an average of the intentions of the fellow denizens 
expressed in a recent past. This average can be seen as a space dependent `cultural' field which encodes all the local, stable features that influence the final decision. 
In essence, this component should be relatively 
smooth over both space and time. The last term describe the influence of the {\it decision} of others, with couplings $J_{ij}$ that measure the strength with which the decision/intention of individual $j$ influences $i$. 
$J > 0$ means conformity of choices, whereas $J < 0$ encodes dissent. Many situations (like the movie and the clapping examples 
given above) are described by a mean-field coupling to the average decision of others: $J_{ij} = J_0/N$, $\forall i \neq j$, or to the average intention of others. 
Finally, the decision rule is $S_i(t)=\Theta(\varphi_i(t)-\Phi_c)$, with $\Theta(x \geq 0)=1$ and $\Theta(x < 0)=0$. \footnote{The process leading from $\varphi$ to $S$ is assumed to be 
deterministic and instantaneous when the final decision is taken. But in fact, one could add an extra source of randomness by assuming that the {\it probability} to choose $S=1$ grows as 
a certain smooth function of $\varphi-\Phi_c$ without changing the essence of the discussion to follow. In fact, this randomisation can be re-absorbed in an appropriate change of 
$P(\epsilon)$.} Note that within such a threshold model, the intention field is defined up 
to an arbitrary scale and shift. There is no physical unit of intention, nor any particular meaning to $\varphi=0$: 
only {\it differences} of intentions can matter in the evolution of the $\varphi_i$'s. 

Note that without the `cultural field' $\phi(\vec R,t)$ and with a random static idiosyncratic term $\epsilon_i$, Eq.~(\ref{conviction}) boils down 
to the Random Field Ising Model~\cite{galam1,sethna_houches}, with first applications to social dynamics appearing in~\cite{galam2,galam3}, see also \cite{collective_shift,nadal_dicrete_choices}. 
For early studies and/or critical reflections in sociophysics~\cite{quetelet}, see e.g. \cite{ostwald,majorana,kadanoff,batty,montroll,toulouse}.

As noted above the intra-{\it commune} correlations between votes may be due to direct imitations between members of the same family (between which intentions are often shared). 
But the long-ranged spatial correlations cannot be due to the imitation of decision term. One reason is, as noted above, that elections are not situations where the actual decision 
of others can matter, since it is known too late. Interestingly, however, the data itself strongly rejects a model where the field $\phi(\vec R,t)$ is 
short-ranged correlated, while assigning the spatial correlations of $\tau$ to a coupling term $J_{\alpha\beta}$ between nearby {\it communes}. 
The reason this model cannot be made to work is the 
following: for long-ranged correlations to emerge, the coupling $J$ must be such that the system is close to its critical point $J_c$, beyond which imitation is 
so strong that the solution of the coupled equations giving the $\{S_i\}$ becomes multi-valued. But when this is the case, the corresponding distribution of turnout rates becomes
very wide, or even bimodal, and negatively skewed in a way that is incompatible with the unimodal, positively skewed and rather narrow distribution observed empirically (see Fig.~\ref{fhisto-abst}). 
A detailed discussion of the inadequacy of this model in the case of elections is provided in \cite{Borghesi}. The long-ranged correlations of $\tau$ should, we believe, be sought in the 
spatial structure of the cultural field $\phi(\vec R,t)$. 

From now on, we therefore neglect the direct imitation component in Eq. (\ref{conviction}) above. We introduce the cumulative distribution of the instantaneous, idiosyncratic component $\epsilon$: ${\cal P}_>(x)=\int_x^\infty dx' P(\epsilon)$. Calling $\pi_\alpha = {\cal P}_>
\left(\Phi_c - \phi(\vec R_\alpha)\right)$ the theoretical turnout rate of {\it commune} $\alpha$, the {\it realized} turnout rate is given by:
\be
p_\alpha=\frac{1}{N_\alpha} \sum_{i=1}^{N_\alpha} S_i \approx \pi_\alpha + \sqrt{\frac{\pi_\alpha(1-\pi_\alpha)}{N_\alpha}} \xi,
\qquad (N_\alpha \gg 1)
\ee
where  $\xi$ a standardised Gaussian noise. We will assume a logistic distribution for $P(\epsilon)$, 
which will make the following discussion particularly transparent, i.e.:
\be
{\cal P}_>(x) = \frac{1}{1+e^{\frac{x-\mu}{\sigma_\epsilon}}},
\ee
where $\mu$ is the average of the $\epsilon$ and $\sigma_\epsilon$ the width of $P(\epsilon)$ that we assume, for
simplicity, to be constant in space and in time, and that we set equal to unity in the following. The average $\mu$ may itself depend both on space and time, and can be seen as an 
extra, {\it commune} specific spatial noise that adds up to the smooth cultural field $\phi$. \footnote{We actually expect $\mu$ to be different in
different neighbourhoods of the same city, reflecting socio-professional or ethnic intra-{\it communes} variability.} 
In fact, we will subtract from $\phi(\vec R)$ any short-range correlated component that we assign to 
$\mu(\vec R)$, such that by definition, the correlation function of the $\phi$ field $C_\phi(r)$ tends smoothly towards $1$ when $r \to 0^+$.

With this particular logistic choice above, one finds:
\bea
& \tau_\alpha=\ln(\frac{p_\alpha}{1-p_\alpha})\\
& \approx \left(\phi(\vec R_\alpha)+\mu(\vec R_\alpha)-\Phi_c\right) + 
\frac{1}{\sqrt{\pi_\alpha(1-\pi_\alpha)N_\alpha}} \, \xi
\eea
Up to a shift and a noise component that vanishes when $N_\alpha \to \infty$, $\tau_\alpha$ and $\phi(\vec R_\alpha)$ are now the very same object. For other choices of the distribution of $\epsilon$, 
such a strict identification is not warranted, but we expect that both object share similar statistical properties. 

Within the above identification, the variance of $\tau_\alpha$ is given by: 
\be\label{sigma}
\sigma^2_\tau = \sigma^2_\phi + \sigma^2_\mu + \langle \frac{1}{Np(1-p)} \rangle \equiv \sigma^2_\phi \left[1+A\right],
\ee
where the ratio of variances $A$ is introduced for later convenience. The variance of the logarithmic turnout rate is therefore the result of three effects: (1) the fluctuations of the smooth cultural field, $\sigma^2_\phi$. This quantity is
not attached to a particular {\it commune} and therefore cannot depend on $N$; (2) the fluctuations of the average intentions in a given {\it commune}, $\sigma^2_\mu$, 
that may well depend on $N$ (a naive guess would be as $N^{-1}$); and (3) the binomial noise effect, that scales as $N^{-1}$. So the extra noise found in the data (see Fig. \ref{ftau-N}), 
that we interpreted above as a sign of intra-{\it commune} herding, may in fact reveal the contribution of $\sigma^2_\mu$. 

A way to distinguish the two interpretations is to compute the {\it covariance} of the $\tau_\alpha$ for different elections as a function of $N$, 
and compare it to the variance of $\tau_\alpha$, plotted in Fig. \ref{ftau-N}. Since the binomial noise is uncorrelated from election to election, its
contribution must drop out from the covariance of $\tau$ for different elections, averaged over all {\it communes}:
\bea
& {\rm{Cov}}_\tau(t,t')=\langle \tau_\alpha(t) \tau_\alpha(t') \rangle_c \\
& = \langle \phi_\alpha(t) \phi_\alpha(t')\rangle_c + \langle \mu_\alpha(t) \mu_\alpha(t') \rangle_c,
\quad (t \neq t'),
\eea
where the index `c' means that we take a cumulant average over all pairs of elections and all {\it communes}.
If the binomial noise was the only source of the $N$ dependence of $\sigma^2_\tau$, the above covariance of the $\tau$s at different times should be independent of $N$. 
This is not what the data shows. There is indeed a residual $N$-dependence that must be ascribed to the statistics of the average of idiosyncratic intentions, $\mu(\vec R)$ 
(although we have no interpretation for the $\sim N^{-3/2}$ dependence that we observe. The assumption that the dispersion of individual biases, $\sigma_\epsilon$, is 
{\it commune}-independent might in fact not be warranted). Interestingly, we have found that the following relation 
accounts very well for the data:
\be
\langle {\rm{Cov}}_\tau(t,t) \rangle_t \approx B \langle {\rm{Cov}}_\tau(t,t') \rangle_{t \neq t'} + \left\langle \frac{1}{Np(1-p)} \right \rangle,
\ee
where $B \approx 1.5$. The remarkable point of this analysis is that the coefficient of the binomial contribution is found to be exactly unity, 
meaning that within this interpretation one does not need to invoke the presence of intra-{\it communes} ``clones'', or more precisely that the probability $z$ of 
herds is too small to be detectable using our data set \footnote{In fact, leaving the
coefficient $B'$ in front of the binomial contribution free, a regression analysis leads to $B' \approx 0.7 < 1$, whereas the presence of 
``clones'' would require $B' > 1$.} The coefficient $B$ simply accounts for the average decay of the correlation of 
$\phi$ and $\mu$ as a function of time, and its value is compatible with the results shown in Fig.~\ref{fcor-temporel} below. 

\section{The random diffusion equation} 
\label{randif}
We will now argue why it is natural to expect that the cultural field $\phi(\vec R_\alpha,t)$ should obey a noisy diffusion equation akin to Eq. (\ref{EW}) with an extra global, time 
dependent driving term. 
Although $\phi(\vec R_\alpha,t)$ is an rather abstract object, the existence of which we postulate, its time evolution should contain a random term that describes events that are specific 
both in time and in space and contribute to changing the overall mood of a given city, such as the closing down of a factory or of a military base, important changes in population, 
or a particularly charismatic local leader, for example. This is captured by the term $\eta(\vec R,t)$, which we imagine to be correlated in space over some length scale $\ell_c$ comparable 
to the typical inter-{\it commune} distance, and over a time $T_c$ of, plausibly, several months.

The Laplacian term describes the fact that people themselves move around and carry with them some components of the local cultural specificity $\phi(\vec R_\alpha,t)$. 
This can be either by actually moving to a nearby city, or by  just visiting or interacting with acquaintances from the neighbouring cities. The all year round exchange of ideas, informations and 
experiences must lead to a local propagation of $\phi(\vec R_\alpha,t)$. Since only difference of intentions matter, the evolution of the cultural field at $\vec R_\alpha$ can only depend on the 
differences $\phi(\vec R_\beta,t)-\phi(\vec R_\alpha,t)$; furthermore, the model must be invariant under a change of scale of $\phi$. In order to satisfy these constraints, 
the most general term describing the evolution of $\phi(\vec R_\alpha,t)$ due to the surrounding influences takes the following, linear form:
\be\label{transfer}
\left.\frac{\partial \phi(\vec R_\alpha,t)}{\partial t}\right|_{\rm{infl.}} = \sum_\beta \Gamma_{\alpha\beta} [\phi(\vec R_\beta,t)-\phi(\vec R_\alpha,t)],
\ee
where $\Gamma_{\alpha\beta}(r_{\alpha\beta}) \geq 0$ is a symmetric influence matrix, that we assume to decrease over a distance corresponding to typical daily displacements of individuals, 
say $10$ km or so. The above equation means that through human interactions, the cultural differences between nearby cities tend to narrow. 
As is well known, the continuum limit of the right hand side of Eq. (\ref{transfer}) reads $\vec \nabla \cdot [ D(\vec R) \vec \nabla \phi(\vec R,t) ]$, with $D(\vec R_\alpha)=\frac12 \sum_{\beta} 
r_{\alpha\beta}^2 \Gamma_{\alpha\beta}$ is a (space dependent) measure of the speed at which the cultural field diffuses. This spatial dependence is a priori expected: for example, $D(\vec R)$ 
should presumably be 
smaller in mountain regions (because of the difficulty to travel from one valley to the next) or in sparsely populated areas (because of the larger distance between neighbouring towns).

Our final equation, that respects all the symmetries of the model \footnote{Note that $\phi \to -\phi$ is another symmetry of the model, 
corresponding to the fact that the propensity to do something is related to minus the
propensity not to do the same thing, since $\Theta(\varphi-\Phi_c)=1-\Theta(\Phi_c-\varphi)$. The scale invariance and the shift symmetry of the 
threshold model in fact only allow higher order 
{\it linear} derivatives to appear, such as $\nabla^4 \phi$, etc. These terms do indeed appear in a gradient expansion of Eq. (\ref{transfer}) above.}, is therefore (in the continuum limit):
\be\label{EW2}
\frac{\partial \phi(\vec R,t)}{\partial t} = \vec \nabla \cdot [ D(\vec R) \vec \nabla \phi(\vec R,t) ] + \eta(\vec R,t) + \nu(\vec R) F(t),
\ee
where $F(t)$ represents the public information, for example the subject and importance of the election, national TV programs, etc. In principle, different {\it communes} may
react differently to the same public information, leading to space dependent `reactivity' $\nu(\vec R)$, but we will neglect these spatial fluctuations in the 
following. \footnote{Short-ranged correlated noise in $\nu(\vec R)$ does not affect the logarithmic behaviour of $C_\phi(r)$ provided $F(t)$ has a finite correlation time.}
Similarly, we assume that the variance and higher cumulants of $\eta$ are homogeneous and independent of the size of the {\it commune}, but this assumption could be relaxed if needed.
Because any average trend is described by $F(t)$, the noise $\eta(\vec R,t)$ is of zero mean. The average trend can be accounted for by a uniform shift of $\phi$:
\be
\phi(\vec R,t) \longrightarrow  \phi(\vec R,t) - \nu  \int_0^t {\rm d}t' F(t').
\ee
This explains why the average value of $\phi$, and therefore of the turnout rate $\tau$, can change substantially over time, while the fluctuations of $\phi$ and its spatial correlations  
remain essentially stable -- as found empirically, see Figs. 1 and 3.

When $D(\vec R)=D$ is uniform, we recover Eq. (\ref{EW}) and the logarithmic decay of the correlation function $C_\tau(r) \propto C_\phi(r)$ follows immediately, since the
contribution of the idiosyncratic $\mu$ field and the binomial noise, both short ranged in space, rapidly disappear when $r > 0$. Since we have removed from $\phi$ its short range 
spatial contribution, the coefficient $\Lambda^2$ appearing in Eq. (\ref{log}) is given by $1/\ln \frac{L}{\ell_c}$. Therefore, 
\be
C_\tau(r \neq 0) = \frac{C_\phi(r)}{1+A} \approx \frac{1}{1+A} \times \frac{\ln \frac{L}{r}}{\ln \frac{L}{\ell_c}}, 
\ee
where $A$ is the ratio of the variance of the idiosyncratic and binomial noise to that of the cultural field, defined in Eq. (\ref{sigma}) above. From Fig. \ref{fcor-spatial}, one
sees that $1/(1+A) \sim 0.2 - 0.4$ for turnout rates: the cultural field explains roughly a third of the variance of the local results. This ratio increases to one half for 
winning votes, except for the 2000 referendum (see Fig. \ref{fcor-spatial}, right plot), meaning that the role of the cultural field is arguably stronger for polarized political decisions than it is for the question of participating or not to a vote.

When $D(\vec R)$ is non uniform, the equilibrium correlation of $\phi(\vec R)$ cannot be computed in general. 
However, drawing analogies from physics \cite{physrep}, we know that this inhomogeneous 
diffusion equation can be, under rather general hypothesis on the statistics of $D(\vec R)$, ``homogeneized''. This means that coarse graining on sufficiently large scales, larger than the 
correlation length of $D(\vec R)$, the effective equation becomes identical to Eq. (\ref{EW}) with an effective diffusion constant $\overline{D}$ that can be computed using e.g. perturbation 
theory, or effective medium approximations \cite{physrep}. What is of interest for our discussion is that the logarithmic dependence of $C_\phi(r)$, Eq. (\ref{log}), is still valid on large enough scales.

%Correl simul
\begin{figure*}
\begin{center}
 \includegraphics[width=8.5cm, height=6cm, clip=true]{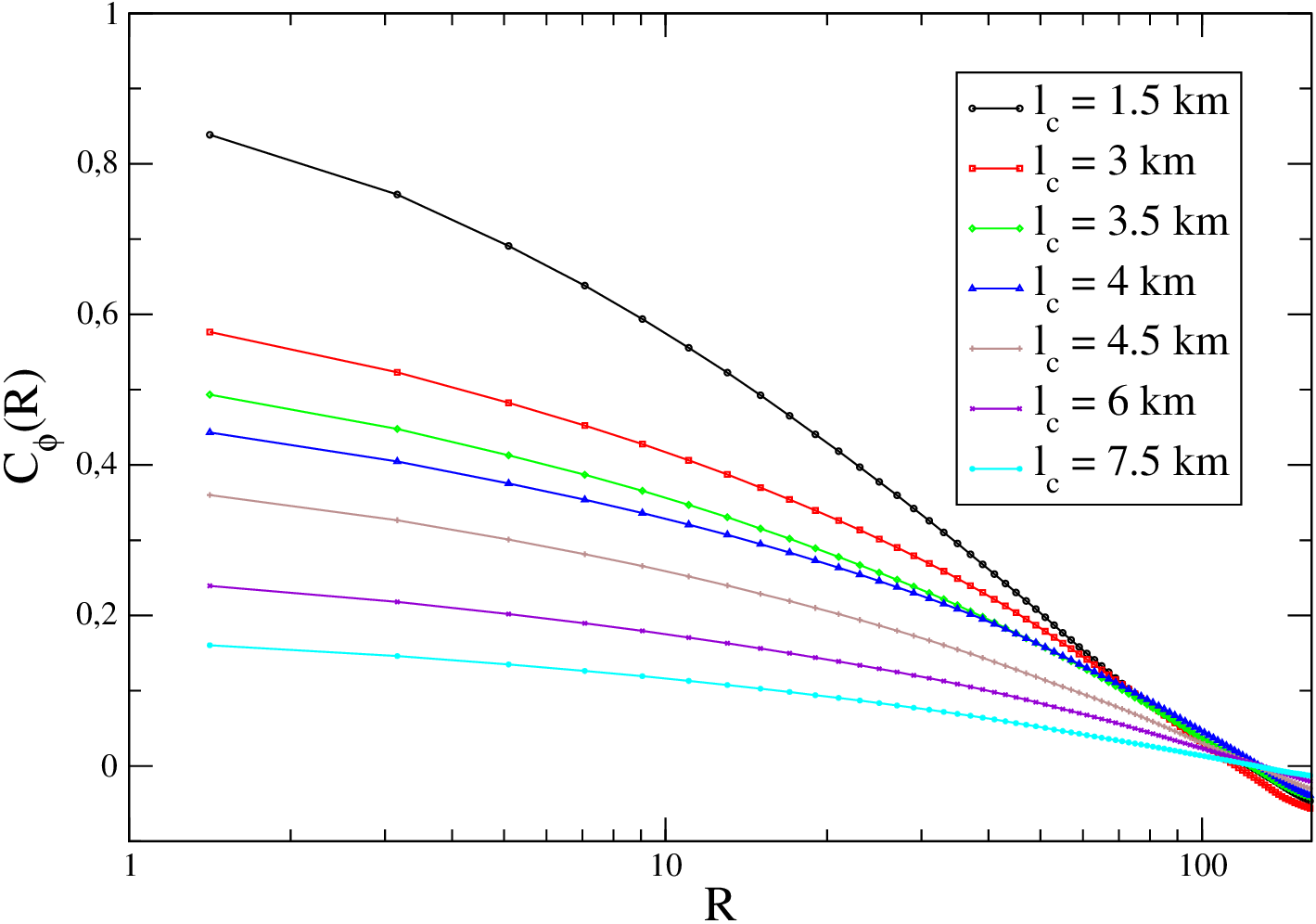}\hfill%, height=6cm
 \includegraphics[width=8.5cm, height=6cm, clip=true]{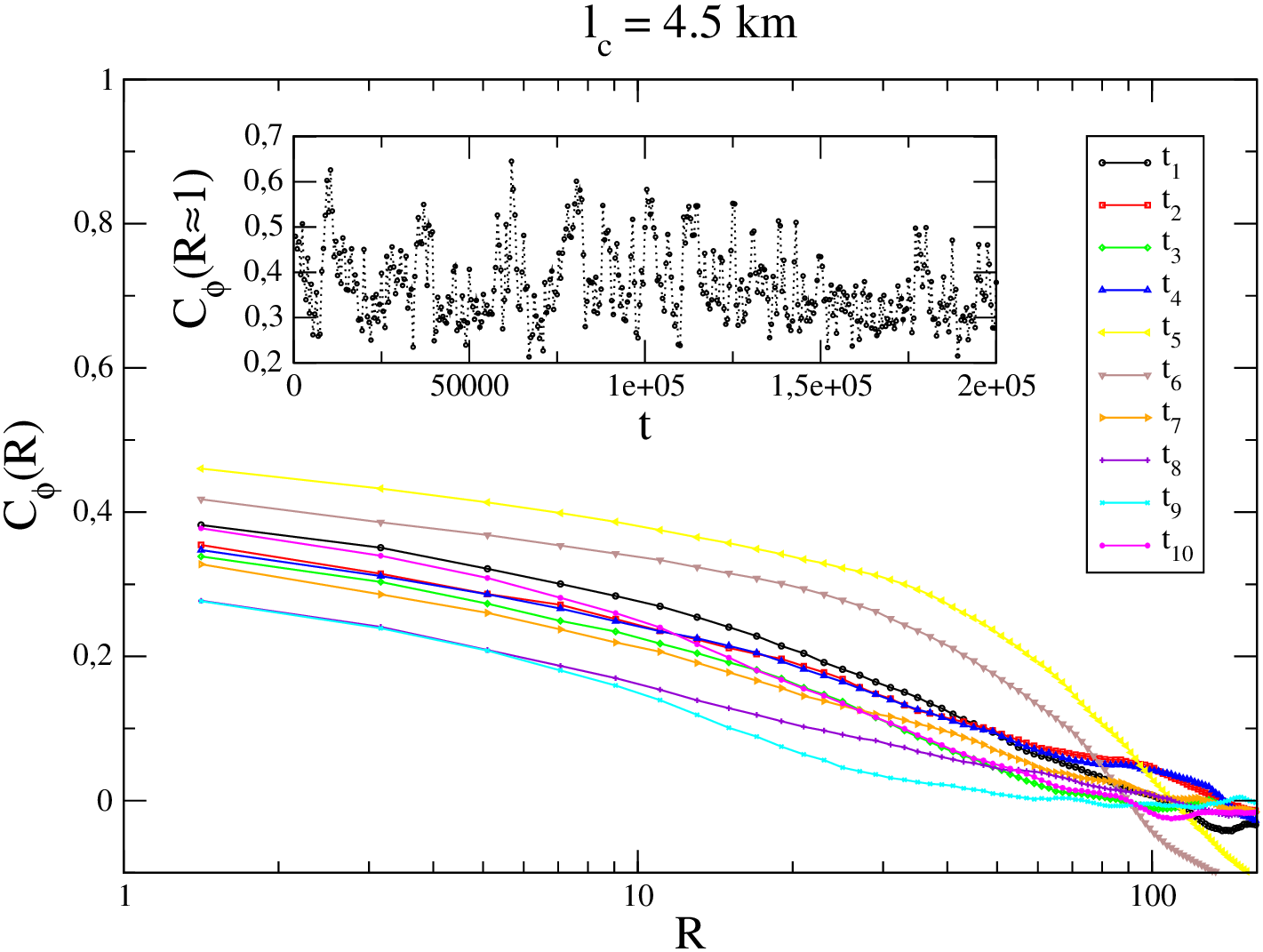}%, height=6cm
\end{center}
\caption{Numerical simulation of the diffusive field model, with the true positions of the French {\it communes}, and an interaction between nearby {\it communes} decaying as 
  $\exp(-r/\ell_c)$. Left, we show the correlation $C_\phi(R)$ as a function of $\ell_c$, after transient regime and averaged over many realizations of the `noise' $\eta$. The distance $R$ is
  here measured in units of the average inter-{\it commune} distance (that depends on the {\it departement}. $R=100$ corresponds to $270$ km, precisely the scale at which the empircal 
  correlations reach zero. Right, ten realizations of $C_\phi(R)$ for $\ell_c=4.5$~km, still after transient regime, as a function of the scaled distance $R$. 
  In the inset, value of the nearest neighbour correlation $C(R=1)$ as a function of `time', showing strong fluctuations.}
% \label{fcor-simul}
\end{figure*}

In order to test these ideas more quantitatively, we have simulated the model using Eq. (\ref{transfer}) with the exact locations of all french {\it communes}, and $\Gamma_{\alpha\beta}(r)=e^{-r/\ell_c}$. 
The spatially inhomogeneities are therefore treated exactly. We show the resulting spatial correlations of $\phi(\vec R)$ in Fig. 4-left, for different values of $\ell_c$, in the range $[1.5,7.5]$ km,
i.e. comparable with the inter-town average distance. We see that $C_\phi(r)$ is indeed approximately logarithmic and looks actually very similar to the empirical curves. Note in particular that
while the small $r$ value of $C_\phi(r)$ and the logarithmic slope $\Lambda^2$  significantly depends on $\ell_c$, as expected, the point $r$ at which $C_\phi(r)$ reaches zero is approximately constant, and, 
remarkably, very close to its empirical counterpart ($L \approx 300$ km, on the order of the size of the system). It is tempting to associate the apparent dependence of $\lambda^2$ on the election 
to a temporal evolution of $\ell_c$. However, this is not correct: our numerical simulations show that even when the system is equilibrated (i.e. when $t \gg L^2/D$), the measured $C_\phi(r)$ for 
a fixed value of $\ell_c$ fluctuates quite strongly between different realizations of the noise $\eta$, see Fig. 4-right. The range over which $\Lambda^2$ varies in the simulation 
is very similar to what is found empirically. 
Therefore, the empirical results are fully compatible with a fixed value of $\ell_c$. Matching the average empirical slope to the numerical data and taking into account the value of $A$ suggests 
that $\ell_c$ must be very small, of the order of a few kms, fully compatible with our interpretation. We have checked in our numerical simulations that our diffusive field model indeed 
reproduces the strong linear correlation between $\tau_\alpha$ and the 
average value of $\tau_\beta$ over the neighbouring towns, with unit slope. This is of course expected and is the essence of the model, since the Laplacian coupling in Eq. (\ref{EW}) says precisely that 
on average the Laplacian of $\phi$ is zero, i.e. that $\phi_\alpha$ is on average equal to $\sum_{\beta \in n(\alpha)} \phi_\beta$, where $n(\alpha)$ is the neighbourhood of $\alpha$. This
empirical finding can in fact be seen as a direct, microscopic motivation for postulating Eq. (\ref{transfer}) above.

\begin{figure}
  \includegraphics[width=8.5cm, height=6cm, clip=true]{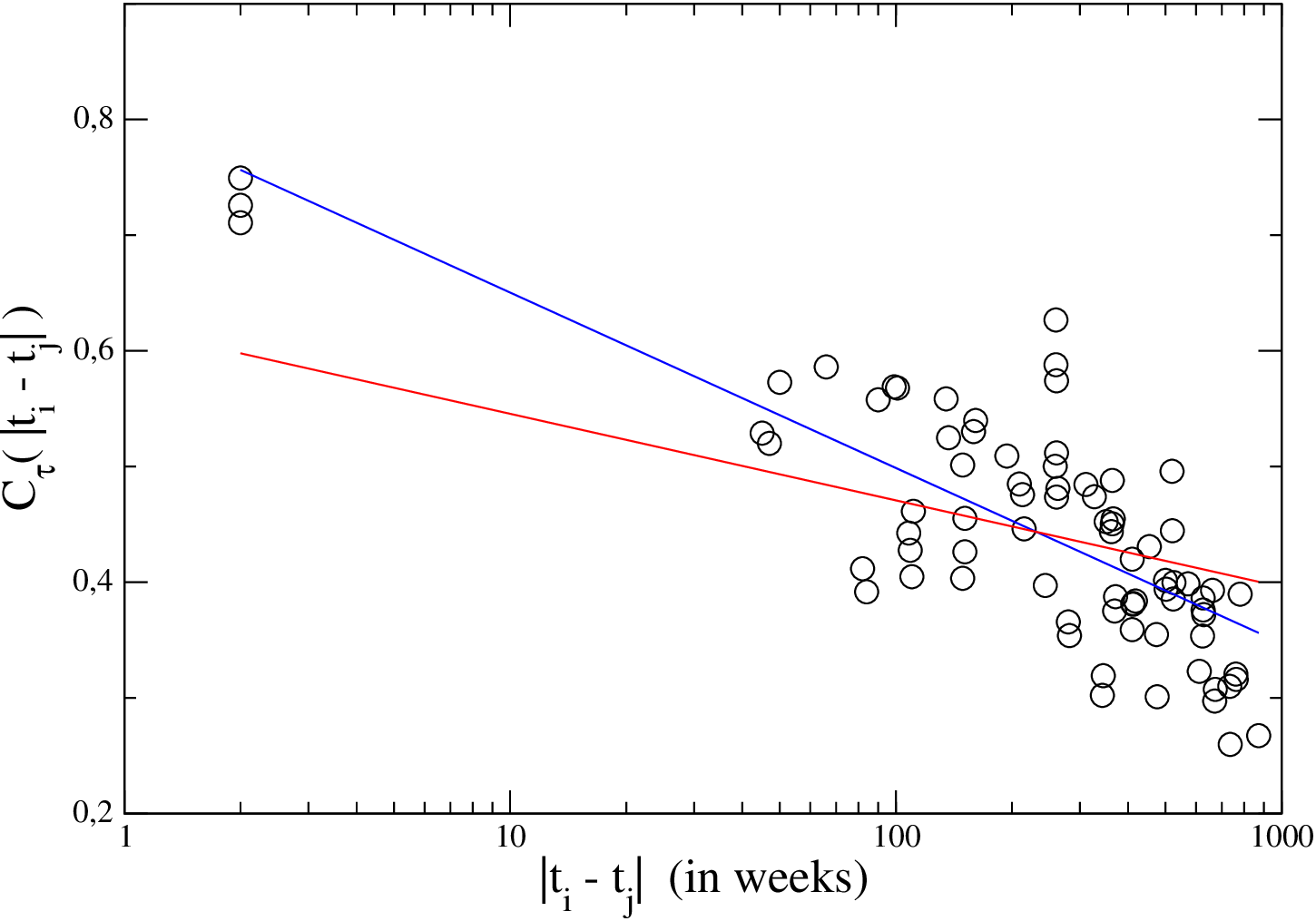}
  \caption{Temporal correlation $C_\tau(t)$ of turnout rates $\tau$. Here, $t=|t_i-t_j|$ where $t_i$ and $t_j$ are the election dates of elections. 
  The red line shows the expected $C_\tau(t)\sim -\frac{1}{2}\lambda^2 \ln(t)$ where $\lambda^2\approx 0.065$ is the averaged slope over all elections of the spatial correlations 
  $C_\tau(r)\sim -\lambda^2 \ln(r)$. The blue line shows the linear regression of $C_\tau(t)$ as a function of $\ln(t)$, here with a slope of $\approx -0.066$.}
 \label{fcor-temporel}
\end{figure}

In order to estimate the diffusion of intentions $D$, one can assume that a substantial fraction of the population of a given town experiences some interaction with neighbouring
towns after several weeks. For opinions to get closer, this time is probably more on the scale of months. Taking the relevant inter-town distance to be $\sim 10$ km leads to a diffusion constant of the 
order of $D \sim 100$ km$^2$/month or less (ways of obtaining more precise empirical estimates of this quantity would be important here). The corresponding equilibration time $T_{eq}$ over $L=300$ km 
is therefore quite long, a century or more. But this is compatible with the fact that the voting habits of the different regions seem to be extremely persistent in history -- which is 
in itself a strong argument for the existence of a cultural field that keeps the memory {\it independently of the presence of particular individuals}. This justifies the apparent correlation 
between geography and votes discussed long ago by Siegfried \cite{Siegfried} ({\it Le granite vote \`a droite, le calcaire vote \`a gauche}).
Interestingly, a more precise prediction of our model is that the temporal autocorrelation of $\phi$ for a given $\vec R$ should behave as: 
\bea
C_\phi(t) = \frac{\left \langle \phi(\vec R,t) \phi(\vec R,0) \right \rangle}{\langle \phi(\vec R,t)^2 \rangle} & \approx - \frac{\Lambda^2}{2} {\ln \frac{t}{T_{eq}}},\\
& \frac{\ell_c^2}{D}, T_c \ll t \ll T_{eq},
\eea
where the slope is exactly one half of the slope of $C_\phi(r)$ versus $-\ln r$, and $T_c$ the correlation time of the noise $\eta$. We have tested this prediction directly on data, by 
studying the autocorrelation of the fluctuations of the participation rate for a given {\it commune} across different elections, averaged over all {\it communes}. Although the data is noisy, 
see Fig.~\ref{fcor-temporel}, the results are indeed compatible with a logarithmic decay in time with a slope that has the correct order of magnitude. 
From the spatial correlations of $\tau$ we find $\lambda^2 \approx 0.065$ (averaged over all elections -- see Fig. \ref{fcor-spatial}), whereas the linear regression of $C_\tau(t)$ as a function of $\ln t$ gives as slope of $\approx -0.066$. The discrepancy with 
the factor 2 predicted by the theory could be related to the fact that the autocorrelation in time is substantially larger than the correlation in space 
(compare Figs. \ref{fcor-spatial} and \ref{fcor-temporel}), meaning that the idiosyncratic field $\mu(\vec R,t)$ has substantially longer temporal correlations than spatial correlations. 
If the correlation time of the noise $\eta$ is of several months, the contribution of the decorrelation of $\mu(\vec R,t)$ to $C_\tau(t)$ probably interferes with the
contribution of the cultural field $\phi$ and effectively increases the empirical slope. In any case, in the absence of more information on the dynamics of the $\mu$ field, 
we find the overall agreement between the model and the behaviour of these temporal autocorrelations satisfying.

%Correl temporelle

\section{Conclusion}

Let us summarise the main messages of this study. First, the statistics of the turnout rates in French elections is found to be surprisingly stable over time, once the average turnout rate is factored in.
The size dependence of the turnout rate variance may have suggested some intra city `herding' effect, but we believe that the data is more consistent with small towns having a larger dispersion in local, idiosyncratic biases. A convincing argument for why this should be so is however lacking. An explanation could be the systematic difference in the socio-cultural background of the population 
of large cities compared to that of small cities. 

Second, for our whole set of elections, the spatial correlation of the turnout rates, or of the fraction of winning votes, is found to decay {\it logarithmically} with the distance between towns. This slow decay of 
the correlations is characteristic of a diffusive random field in two dimensions. This result is robust against many non essential modifications of the basic version of the model, much as the statistics of a random walk is robust against modifications of the microscopic 
construction rules. Based on these empirical observations and on the analogy with the two-dimensional random diffusion equation, we have proposed that individual decisions can be
rationalised in terms of an underlying ``cultural'' field that locally biases the decision of the population of a given region, on top of an idiosyncratic, city-dependent field, with short 
range correlations. 

Based on symmetry considerations and a set of plausible arguments, we have suggested that this cultural field obeys an equation in the universality class of the random diffusion equation, Eq. (\ref{EW2}) above. We believe that similar considerations should hold for other decision processes, such as consumption habits, behavioral biases, etc. More empirical work on the spatial correlations of these 
decisions, in different situations and in different countries, would be very valuable to test our claim of universality. 
Direct estimates of the parameters of the model, such as the value of 
the diffusion constant $D$ or the relative strength of the idiosyncratic field, are clearly needed at this stage. We hope that our work will motivate more empirical studies to refine and calibrate the model proposed here.

\section*{Acknowledgements}

C. B. would like to thank Brigitte Hazart, from the \textit{Minist\`ere de l'Int\'erieur, bureau des \'elections et des \'etudes politiques}, for the great work she did to gather 
and make available the electoral data that we used. We also thank J. Chiche, S. Franz, S. Fortunato, J.-P. Nadal and M. Marsili for useful comments.

\end{document}